\documentclass[prd,aps,twocolumn,nofootinbib,preprintnumbers,superscriptaddress,preprintnumbers,balancelastpage,longbibliography]{revtex4-2}

\usepackage{graphicx}
\usepackage{bm}
\usepackage{hyperref}
\usepackage{hyperref}
\usepackage{slashed}
\usepackage{aas_macros}
\usepackage{float}
\usepackage{slashed}
\usepackage{lipsum}
\usepackage{subfigure}
\usepackage{multirow}
\usepackage{amsmath}
\usepackage{array} 
\usepackage{varwidth} 
\usepackage{tabularx}
\usepackage{braket}
\usepackage[dvipsnames]{xcolor}

\usepackage[normalem]{ulem}

\newcommand{\be}{\begin{equation}}
\newcommand{\ee}{\end{equation}}
\newcommand{\bea}{\begin{eqnarray}}
\newcommand{\eea}{\end{eqnarray}}

\hypersetup{
     colorlinks   = true,
     citecolor    = NavyBlue,
     urlcolor     = NavyBlue,
     linkcolor    = NavyBlue
}

\usepackage{listings}
\usepackage{color,xcolor}

\begin{document}



\title{Super-Kamiokande Strongly Constrains Leptophilic Dark Matter Capture in the Sun}

\author{Thong T.Q. Nguyen}
\thanks{{\scriptsize Email}: \href{mailto:thong.nguyen@fysik.su.se}{thong.nguyen@fysik.su.se}; \href{https://orcid.org/0000-0002-8460-0219}{0000-0002-8460-0219}}
\affiliation{Stockholm University and The Oskar Klein Centre for Cosmoparticle Physics, Alba Nova, 10691 Stockholm, Sweden}

\author{Tim Linden}
\thanks{{\scriptsize Email}: \href{mailto:linden@fysik.su.se}{linden@fysik.su.se};  \href{http://orcid.org/0000-0001-9888-0971}{0000-0001-9888-0971}}
\affiliation{Stockholm University and The Oskar Klein Centre for Cosmoparticle Physics, Alba Nova, 10691 Stockholm, Sweden}

\author{Pierluca Carenza}
\thanks{{\scriptsize Email}: \href{mailto:pierluca.carenza@fysik.su.se}{pierluca.carenza@fysik.su.se};  \href{https://orcid.org/0000-0002-8410-0345}{0000-0002-8410-0345}}
\affiliation{Stockholm University and The Oskar Klein Centre for Cosmoparticle Physics, Alba Nova, 10691 Stockholm, Sweden}

\author{Axel Widmark}
\thanks{{\scriptsize Email}: \href{axel.widmark@fysik.su.se}{axel.widmark@fysik.su.se};  \href{https://orcid.org/0000-0001-5686-3743}{0000-0001-5686-3743}}
\affiliation{Stockholm University and The Oskar Klein Centre for Cosmoparticle Physics, Alba Nova, 10691 Stockholm, Sweden}
\affiliation{Columbia University, 116th and Broadway, New York, NY 10027 USA}

\begin{abstract}
\noindent The Sun can efficiently capture leptophilic dark matter that scatters with free electrons. If this dark matter subsequently annihilates into leptonic states, it can produce a detectable neutrino flux. Using 10~years of Super-Kamiokande observations, we set constraints on the dark-matter/electron scattering cross-section that exceed terrestrial direct detection searches by more than an order of magnitude for dark matter masses below 100~GeV, and reach cross-sections as low as $\sim$~4~$\times$10$^{-41}$~cm$^{-2}$.
\end{abstract}

\maketitle


\noindent\textbf{\emph{Introduction.}} --- Detecting the particle interactions of dark matter is a cornerstone in our efforts to study beyond the standard model physics~\cite{Bertone:2004pz, Bertone:2016nfn, Cirelli:2024ssz}. Many of the most sensitive constraints depend on searching for rare scattering interactions between the dark matter particle and standard model particles~\cite{Kahn:2021ttr, Knapen:2021run, Lin:2019uvt, Blanco:2019lrf, Essig:2022dfa, XENON:2024wpa, LZ:2024trd, PandaX:2022xqx}.

Two disparate strategies motivate current searches. The most popular, used in terrestrial detectors, uses underground facilities to avoid astrophysical backgrounds and provide sensitivity to single dark matter interactions. More recently, there has been an effort to ``go big", using the large mass of celestial objects to constrain rare dark matter interactions in spite of the large background. While some searches focus only on the scattering of dark matter ({\it e.g.}, neutron star heating or asymmetric dark matter interactions~\cite{Bramante:2014zca, Baryakhtar:2017dbj}), most celestial probes focus on particular annihilation properties that allow the dark matter signal to escape from the celestial object.

\begin{figure}[t]
    \centering
    \includegraphics[width=0.98\linewidth]{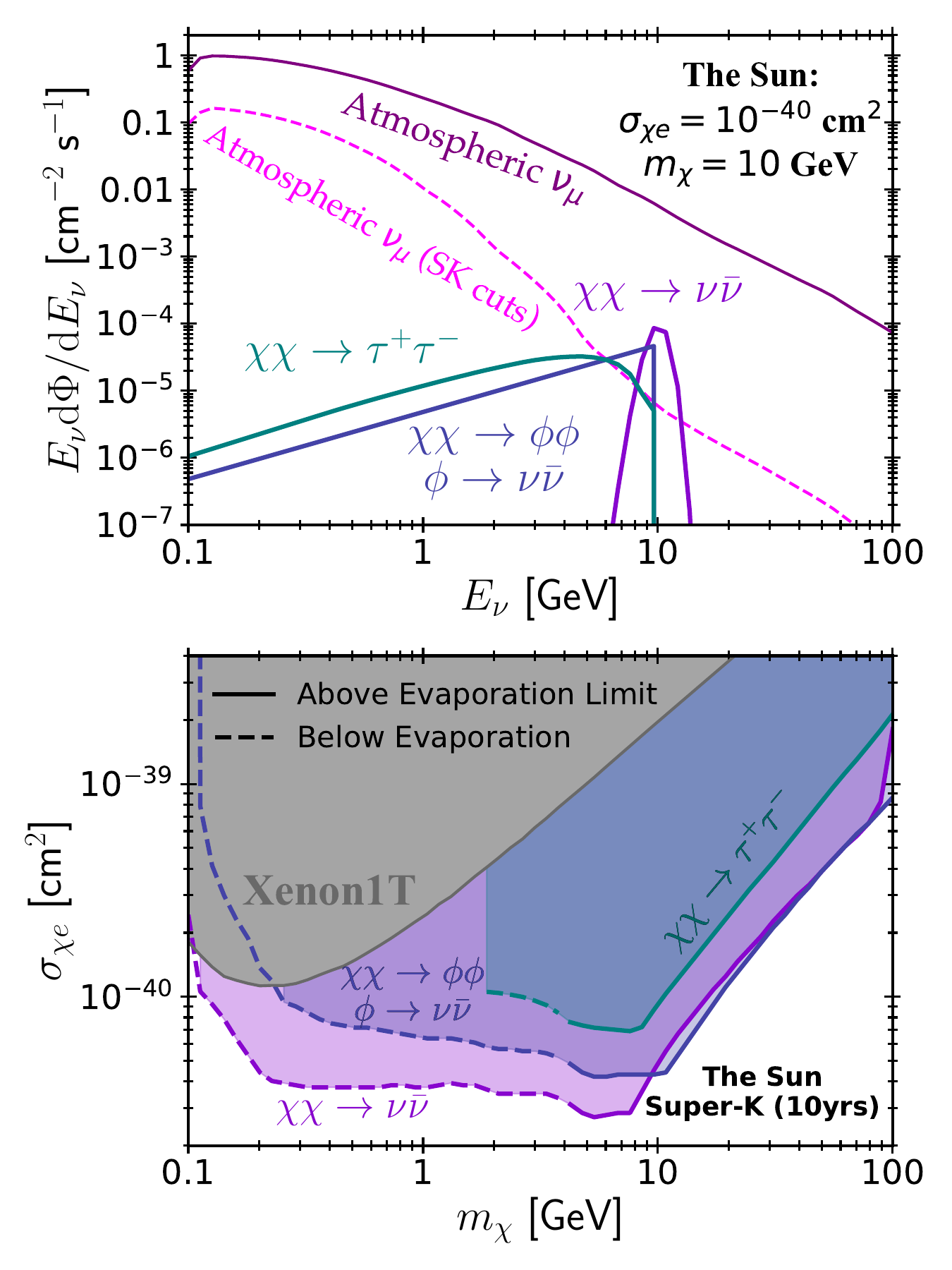}
    \vspace{-0.5cm}
    \caption{({\bf Top}:) Expected neutrino signal from dark matter annihilation in the Sun compared to the atmospheric muon background in Super-K. Results are shown for annihilation to neutrinos (purple, smeared by a 10\% detector resolution), $\tau^+\tau^-$ pairs (teal) and light mediators that decay to neutrinos (blue). ({\bf Bottom}): Super-K constraints on the dark matter-electron cross section, compared to Xenon1T constraints (gray). Dashed lines represent dark matter masses below 4~GeV, where solar evaporation may be important.} 
    \vspace{-0.6cm}
    \label{fig:firstpage}
\end{figure}

To date, most terrestrial and celestial searches have focused on dark matter scattering with baryons, motivated by the similar masses of standard WIMPs and nuclei as well as the large enhancement in the interaction rate for spin-independent nuclear scattering. However, leptophilic dark matter models also provide well-motivated extensions of the standard model, which may be linked to several current anomalies in the neutrino and lepton sectors~\cite{Fox:2008kb, Bertone:2008xr, John:2023ulx, John:2021ugy, Borah:2024twm, Barman:2021hhg, Nardi:2008ix}. Leptophilic models for dark matter \emph{scattering} motivate searches for the subsequent leptophilic \emph{annihilation} of those same dark matter particles. Leptophilic annihilations can be highly advantageous for celestial body searches due to their high neutrino yield, potentially including direct annihilations to neutrino pairs.

The Sun is the preeminent astrophysical body for dark matter searches due to its proximity, which serves to both increase the neutrino flux observed at Earth, and also improve our background rejection, due to its motion across the sky~\cite{Acevedo:2020gro, Berlin:2024lwe, Ng:1986qt, Chauhan:2023zuf, Chu:2024gpe, Garani:2017jcj, Maity:2023rez, HAWC:2022khj, Nisa:2019mpb, HAWC:2018szf, IceCube:2021xzo, Leane:2017vag, Bell:2021esh, Bell:2011sn, Bell:2021pyy, Bell:2012dk, Niblaeus:2019gjk, Edsjo:2017kjk, Bose:2021cou, IceCube:2016yoy, Feng:2016ijc,Kouvaris:2007ay, Kouvaris:2015nsa, Kouvaris:2016ltf, Widmark:2017yvd, Catena:2016ckl, Nguyen:2026nhe}. Solar neutrino observations have been used to probe novel physics, including constraining the dark matter/baryon cross-section across a wide range of dark matter masses.

In this \emph{letter}, we use 10~years of Super-Kamiokande (Super-K) observation to probe leptophilic dark matter capture and annihilation in the Sun. Figure~\ref{fig:firstpage} presents our main result, showing that current solar data rules out electron scattering cross-sections above $\sim$10$^{-40}$~cm$^{-2}$ for dark matter masses below 20~GeV across multiple leptonic annihilation channels. These constraints exceed the sensitivity of terrestrial experiments by more than an order of magnitude, and provide world-leading constraints on leptophilic dark matter scattering.

\noindent \textbf{\emph{Leptophilic Dark Matter Capture }}--- 
Leptophilic dark matter can scatter with electrons inside the Sun and lose kinetic energy. If the dark matter kinetic energy falls below the Sun's escape velocity, it will be captured. In the limit where dark matter particles only scatter with the Sun once\footnote{In App.~\ref{subappen:geo_sig}, we discuss the modifications in Refs.~\cite{Bottino:2002pd, Bernal:2012qh, Leane:2021ihh, Leane:2023woh, John:2024thz, John:2023knt, Ilie:2021iyh, Ilie:2024sos, Cappiello:2023hza} needed when multiple scatterings become non-negligible. However, the results shown in this work do not approach this limit.}, the dark matter particles will be captured at a rate given by~\cite{Garani:2017jcj}: 

\begin{align}
    C^{\rm weak}_{\star}=&\int_{0}^{R_{\star}}{\rm d}r4\pi r^{2}\int_{0}^{\infty}{\rm d}u_{\chi}\Big{(}\frac{\rho_{\chi}}{m_{\chi}}\Big{)}\frac{f_{v_{\odot}}(u_{\chi})}{u_{\chi}}\nonumber\\
    &\times\int_{0}^{v_{\rm esc}(r)}{\rm d}v\times w(r) R^{-}_{e}(w\to v),
    \label{eq:captweak}
\end{align}

\noindent where $R^{-}_{e}(w\to v)$ is the scattering rate that decreases the velocity from $w(r)$ to $v$, which falls below the escape velocity $v_{\rm esc}(r)$.  We set the solar system dark matter density to $\rho_{\chi}=0.3$~GeV/cm$^{3}$, and use the standard velocity distribution $f_{v_{\odot}}$ in the solar system. We provide full results for our scattering rate in App.~\ref{appen:capt_rate}, which primarily follows previous literature.

The scattering rate $R^{-}_{e}$ can be calculated in three different scenarios depending on the interaction between dark matter and the electron~\cite{Garani:2017jcj}: (1) an isotropic and velocity-independent (constant) cross section. (2) an isotropic and velocity-dependent ($v_{\rm rel}^{2}$) cross section, or (3) a transfer momentum-dependent ($q^{2}$) cross section. We primarily show results for scenario 1, but compare each scenario in Figure~\ref{fig:othercapture}.

We note that our capture rate exceeds that of Ref.~\cite{Garani:2017jcj} by a factor of $\sim$3 at low dark matter masses, and up to a factor of $\sim$7 at high dark matter masses, a fact first pointed out in Ref.~\cite{Krishna:2025ncv}, who obtain a smaller dark matter capture rate than in our work. In discussions with the authors of Ref.~\cite{Krishna:2025ncv}, we found that the difference between our results stems entirely from the choice in Ref.~\cite{Krishna:2025ncv} to insert a kinematic cutoff into the formalism for the dark matter/electron scattering cross-section, which is valid only in the case that the electrons have zero temperature. If this cutoff is removed from the calculations of Ref.~\cite{Krishna:2025ncv}, they obtain results identical to ours. The initial kinematic constraint stems from Gould (1987)~\cite{Gould:1987ir} who focused on nuclear scattering in the Earth, a scenario in which making a zero-temperature approximation is reasonable. However, applying such a cutoff (or similar variants of it) to high-temperature electrons artificially suppresses the dark matter capture rate by ignoring the extra stopping power of high-energy electrons that collide with the dark matter head on. We discuss this effect in further detail in the Supplemental Material.

\vspace{0.5cm}
\noindent \textbf{\emph{Neutrinos from Dark Matter Annihilation}} --- Dark matter capture by celestial bodies can produce regions inside the object with large dark matter densities. Dark matter in these regions can efficiently annihilate to standard model particles. In App.~\ref{appen:annihilationrate} we review the equations that govern annihilation inside a celestial body. However, when dark matter annihilation and capture is in equilibrium, the result simplifies to:

\begin{equation}
    \Gamma_{\chi}=\frac{C_{\star}}{2},
    \label{eq:GammaChi}
\end{equation}

\noindent where $\Gamma_{\chi}$ is the annihilation rate. The factor of two arises because each annihilation removes two dark matter particles. Unexpectedly, this shows that annihilation in celestial bodies depends on the dark matter scattering cross section and competes with terrestrial direct detection.

In most cases, the standard model annihilation products will be re-absorbed and thermalized by stellar material. However, neutrinos can travel unimpeded to Earth, providing direct access to the annihilation physics. In this paper, we consider three leptophilic dark matter models that can produce a large neutrino flux: (1) annihilation directly to neutrino final states, (2) annihilation to $\tau^+\tau^-$, which quickly decay to states that include neutrinos~\cite{Chauhan:2023zuf, Maity:2023rez}, (3) annihilation to long-lived mediators that subsequently decay to neutrinos~\cite{Nguyen:2022zwb}.

The first and third cases produce the brightest signal, since 100\% of the dark matter mass is converted into neutrinos. In the second case the neutrinos are secondary products of $\tau$-decay, and only $\sim$8\% of the dark matter mass is converted into neutrinos. However, in counting experiments such as Super-K, the multiplicity of neutrinos from $\tau$ decays can produce comparable limits. 

\begin{figure*}[t]
\centering
\includegraphics[width=1.98\columnwidth]{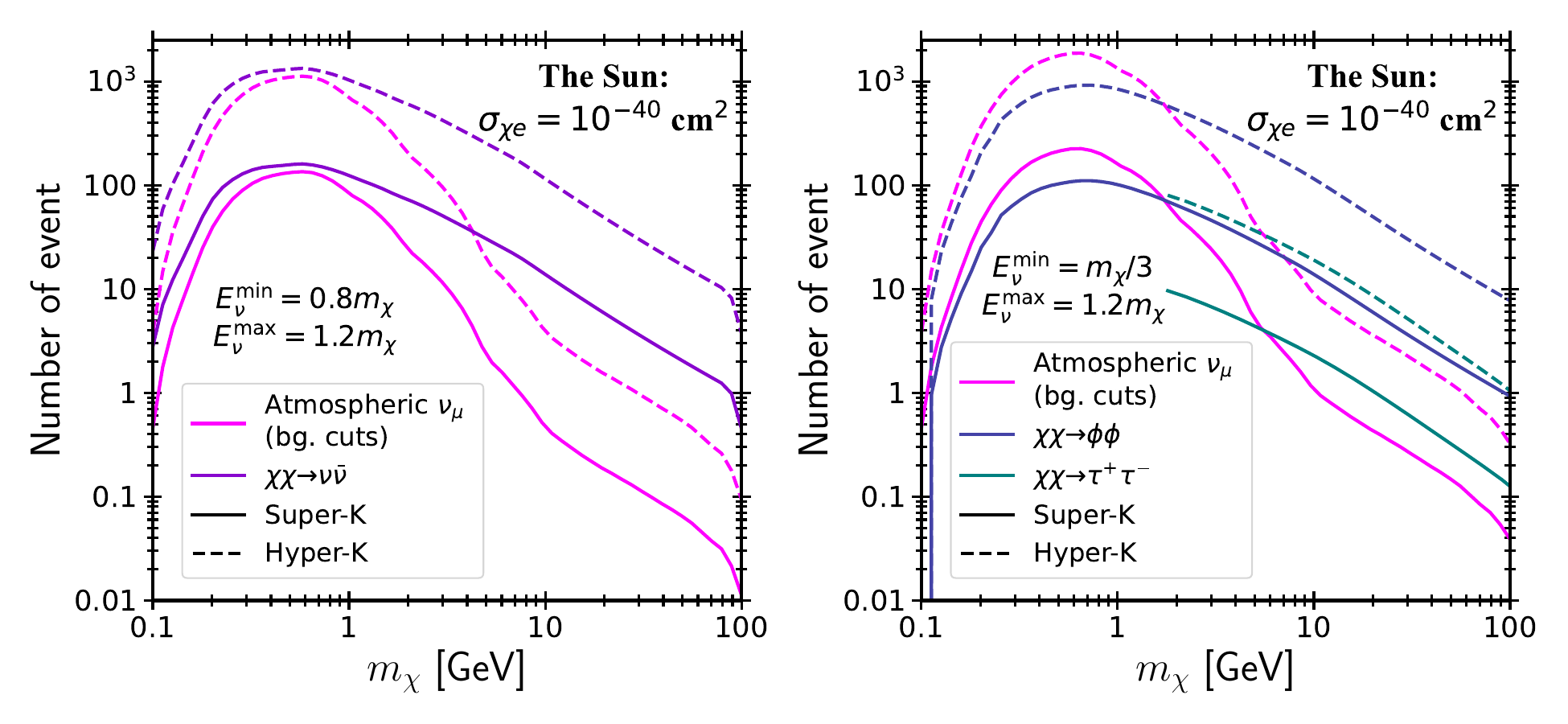}
\vspace{-0.4cm}
\caption{The number of solar dark matter neutrino events for 10~yr exposures with Super-K (solid) and Hyper-K (dashed), assuming $\sigma_{\chi e}=10^{-40}$ cm$^{2}$. ({\bf Left}) $\chi\chi\to\nu\bar{\nu}$ channel (purple) with a 10\% energy smearing at the dark matter mass. ({\bf Right}) $\chi\chi\to\phi\phi$ (blue) and $\chi\chi\to\tau^{+}\tau^{-}$ (teal) integrated over an energy range spanning from  $E_{\rm min}=m_{\chi}/3$ up to 20\% above the dark matter mass. The atmospheric background (magenta) is shown in each case integrated over the same energy range.}
\label{fig:Nevent}
\end{figure*}

The neutrino flux at Earth from leptophilic dark matter annihilation is given by: 

\begin{equation}
E\frac{{\rm d}\Phi_{\nu}}{{\rm d}E_{\nu}}=\frac{\Gamma_{\chi}}{4\pi D^{2}}\times E_{\nu}\frac{{\rm d}N_{\nu}}{{\rm d}E_{\nu}}\times P_{\rm surv},
\end{equation}
where $D$ is the average distance to Earth. This equation represents the total neutrino flux and assumes that neutrino oscillations randomize the neutrino flavor before reaching Earth. However, we only consider muon neutrinos in this work, due to the relatively good angular reconstruction of muon tracks. We note that water Cherenkov detectors are also unable to distinguish muons from anti-muons. Thus, the detectable muon/anti-muon neutrino spectrum includes a factor of 2/3 to account for the flavor and particle/antiparticle multiplicity as

\begin{equation}
    \frac{{\rm d}N_{\nu}}{{\rm d}E_{\nu}}\Big{|}_{\chi\chi\to\nu_{\mu}\bar{\nu}_{\mu}}=\frac{2}{3}\delta(E_{\nu}-m_{\chi}),
\end{equation}
which is a line spectrum that is usually smeared out by the detector energy resolution. For long-lived mediator scenarios, the spectrum is a box distribution given by:
\begin{align}
     \frac{{\rm d}N_{\nu}}{{\rm d}E_{\nu}}\Big{|}_{\phi\to\nu_{\mu}\bar{\nu}_{\mu}}&=\frac{4}{3\sqrt{m_{\chi}^{2}-m_{\phi^{2}}}}\Theta(E-E_{\rm min})\Theta(E_{\rm max}-E)\nonumber\\
     &\simeq \frac{4}{3m_{\chi}}\Theta(m_{\chi}-E),
\end{align}
which assumes that the mediator mass, $m_{\phi}$, is much lighter than the dark matter mass, $m_{\chi}$, and the mediator only decays to neutrinos~\cite{Nguyen:2022zwb}. In this case, the neutrino production probability is given by:
\begin{equation}
    P_{\rm surv}\Big{|}_{\phi\to \nu\bar{\nu}}=1-e^{-D/L},
\end{equation}
where $L$ is the decay length of a mediator with energy $E_{\phi}\simeq m_{\chi}$. We note that this survival probability differs from the standard long-lived mediator case (e.g.,~\cite{Leane:2024bvh}), which is calculated when the mediator decays to electromagnetic states. In that case, the electromagnetic particles are absorbed by solar material if the mediator decays in the Sun, setting the survival probability to 0. Neutrinos, however, also usually escape from the Sun, removing this lower limit. Thus, long-lived mediators can produce a bright neutrino flux as long as they are produced on-shell, even if their decay length is short.

For the case of $\tau^{+}\tau^{-}$ final states, we use PPPC4DM$\nu$~\cite{Cirelli:2010xx, Baratella:2013fya}, which calculates the Solar neutrino spectrum, taking into account that muons and pions which are produced inside the Sun cool before decaying to neutrinos and do not contribute to our $>$100~MeV neutrino spectrum~\cite{Rott:2012qb}. We generate results for dark matter masses between 1.8~GeV and 1~TeV, taking into account electroweak and QCD corrections. We note that other packages can produce similar results~\cite{Arina:2023eic, Bauer:2020jay, Liu:2020ckq}.

Notably, leptophilic dark matter can also annihilate into e$^{+}$e$^{-}$ or $\mu^{+}\mu^{-}$, which can produce neutrinos through weak bremsstrahlung~\cite{Maity:2023rez} or muon decay. However, we do not consider these in our analysis. In the case of muon decay, muon cooling will produce a neutrino spectrum that falls below the 100~MeV cutoff of our analysis~\cite{Rott:2012qb}. In the case of weak bremsstrahlung, the process is suppressed because the 0.1--100~GeV dark matter mass range that we consider mostly lies below the weak scale.

Finally, while most neutrinos escape the solar interior, some can be absorbed by Hydrogen. In this work, we take this attenuation into account, which decreases the survival probability at higher neutrino energies. At $E_{\nu}=100$~GeV, this probability is $\simeq 96\%$, meaning that attenuation has a very small effect on our results through most of the relevant parameter space. We discuss the calculation for this probability in the Appendix~\ref{appen:neutrino_int}.


\begin{figure*}[t]
\centering
\includegraphics[width=2\columnwidth]{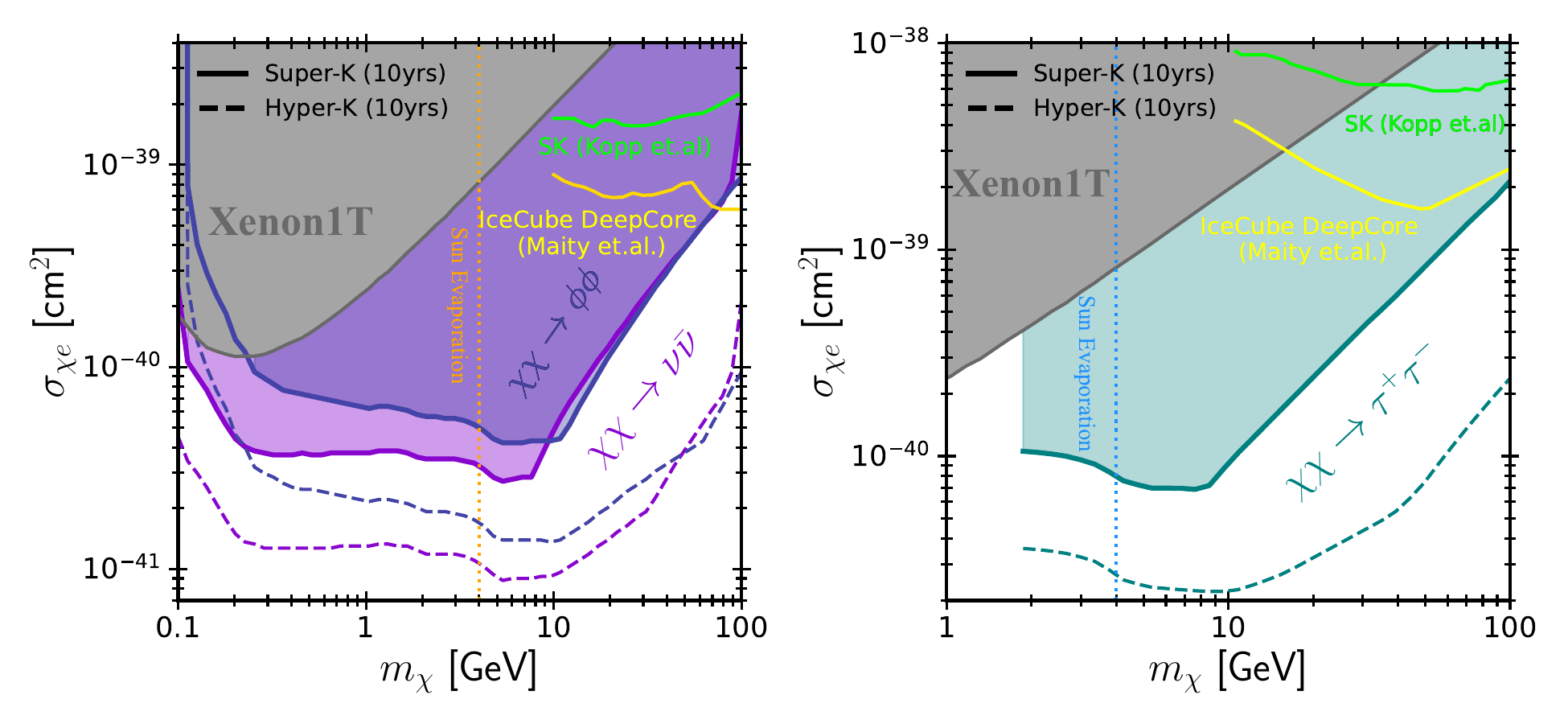}
\caption{Our Super-K constraints and Hyper-K projections compared with previous work, Super-K (lime) data down to 10~GeV and the full-sky atmospheric background~\cite{Kopp:2009et}, and DeepCore (yellow)~\cite{Maity:2023rez}. Results are shown for ({\bf Left}) The $\chi\chi\to\nu\bar{\nu}$ and $\chi\chi\to\phi\phi$ channels and ({\bf Right}) the $\chi\chi\to \tau^{+}\tau^{-}$ channel.}
\label{fig:compare}
\end{figure*}

\vspace{0.5cm}
\noindent \textbf{\emph{Super-K and Hyper-K Observations }}--- Super-K serves as one of the most sensitive detectors for neutrinos with energies between 0.1--100~GeV. Super-K has a fiducial mass of $22.5$~kton of water. The detector has an energy resolution of $\leq$10\%, an angular resolution that varies from $\sim$90$^{\circ}$--2$^{\circ}$ for 0.1--100~GeV neutrino, and is capable of full-sky observations. These features make it an optimal detector for solar system sources~\cite{Shiozawa:1999sd, Drakopoulou:2017apf, Lin:2022jyv,Pospelov:2023mlz, Ema:2024oce, Super-Kamiokande:2019gzr}. The detector is sensitive to muon neutrinos, with an efficiency of $\simeq 80\%$ for charged current (CC) interactions~\cite{Super-Kamiokande:2015qek}.

Figure~\ref{fig:firstpage} (top) shows the full-sky atmospheric muon neutrino spectrum between 0.1--100~GeV (maroon), as calculated by the Super-K collaboration~\cite{Super-Kamiokande:2015qek}. In order to determine the atmospheric neutrino background that could be confused with Solar neutrinos, we apply a cut based on the energy-dependent neutrino angular resolution of a water Cherenkov detector~\cite{Konishi:2010mv, Konishi:2011sc, Galkin:2008qe}, which is a combination of two energy-dependent terms: an irreducible angular uncertainty correlating to the angular distribution of muons that are produced through muon neutrino scattering, and a detector uncertainty that depends on the ability of Super-K to reconstruct muon tracks. The combined angular resolution model is described in App.~\ref{subappen:neutrino_angle}, and is more accurate than previous studies~\cite{Robles:2024tdh}, which assumed a constant angular resolution of ~$20^{\circ}$, independent of neutrino energy.

We calculate the number of observed neutrino events in the energy range $[E_{\nu}^{\rm min}, E_{\nu}^{\rm max}]$ as:
\begin{align}
    N_{\nu}=\int_{E_{\nu}^{\rm min}}^{E_{\nu}^{\rm max}}{\rm d}E_{\nu}\frac{{\rm d}\Phi_{\nu}}{{\rm d}E_{\nu}}\sigma_{\nu H_{2}0}(E_{\nu})\times {\rm Exposure},
    \label{eq:Nevent}
\end{align}

\noindent using the neutrino and water molecule cross section from Ref.~\cite{Zhou:2023mou}, which we discuss in Appendix~\ref{subappen:neutrino_sigma}. For the exposure of Super-K, we assume 22.5~kton of pure water with 10-year time scale. We also project Hyper-Kamiokande (Hyper-K) results using the same 10~yr timescale and a 187~kton mass.

We count the number of neutrino events for both dark matter production in the Sun and from the atmospheric muon neutrino background in different neutrino energy ranges. We calculate the total number of background counts depending on the dark matter mass and final state. For direct annihilation to $\nu\bar{\nu}$, the neutrino flux produces a line-like signal, and we integrate the atmospheric background from 0.8m$_\chi$ to 1.2m$_\chi$, which corresponds to twice the detector energy resolution, allowing us to model the background over a region that effectively captures the entire signal. For annihilation to $\tau^+\tau^-$ and light-mediators, we keep the same upper limit, but adopt a lower-limit of 0.33m$_\chi$ to encompass the region where the majority of the dark matter signal resides. In this case, we ignore the detector energy resolution, which is small compared to the instrinsic smearing of the signal. For consistency, we also ignore dark matter neutrinos below 0.33m$_\chi$ in these cases. We show the results for a benchmark cross section at $10^{-40}$~cm$^{2}$ in Figure~\ref{fig:Nevent}. 

For each dark matter-electron cross section, we calculate the number of muon neutrinos from both signal and background, using the cross sections for muon (anti-)neutrinos with water molecules from Figure~\ref{fig:H2O}. We assume the events are Poisson distributed, and set the 95\% CL on the dark matter neutrino flux such that:

\begin{align}
    \sum_{k=N_{\rm DM}+N_{\rm bg}}^{\infty}\frac{\lambda^{k}e^{-\lambda}}{\Gamma(k+1)}\leq 0.05,
    \label{eq:95CL}
\end{align}

\noindent where $\lambda\equiv N_{\rm bg}(m_{\chi})$ is the expected number of atmospheric muon neutrino events over an energy range consistent with the bounds described above. The null detection of neutrinos from the Solar location lets us a set a background on the expected flux at $\lambda=2.7$. We also notice that there is the Solar Atmospheric neutrino background (SA$\nu$) coming from high energy cosmic rays scattering with the solar atmosphere~\cite{Ng:2017aur, Edsjo:2017kjk}. However, this background is very small in Super-K and does not significant impact our results.

\begin{figure*}[t]
\centering
\includegraphics[width=2\columnwidth]{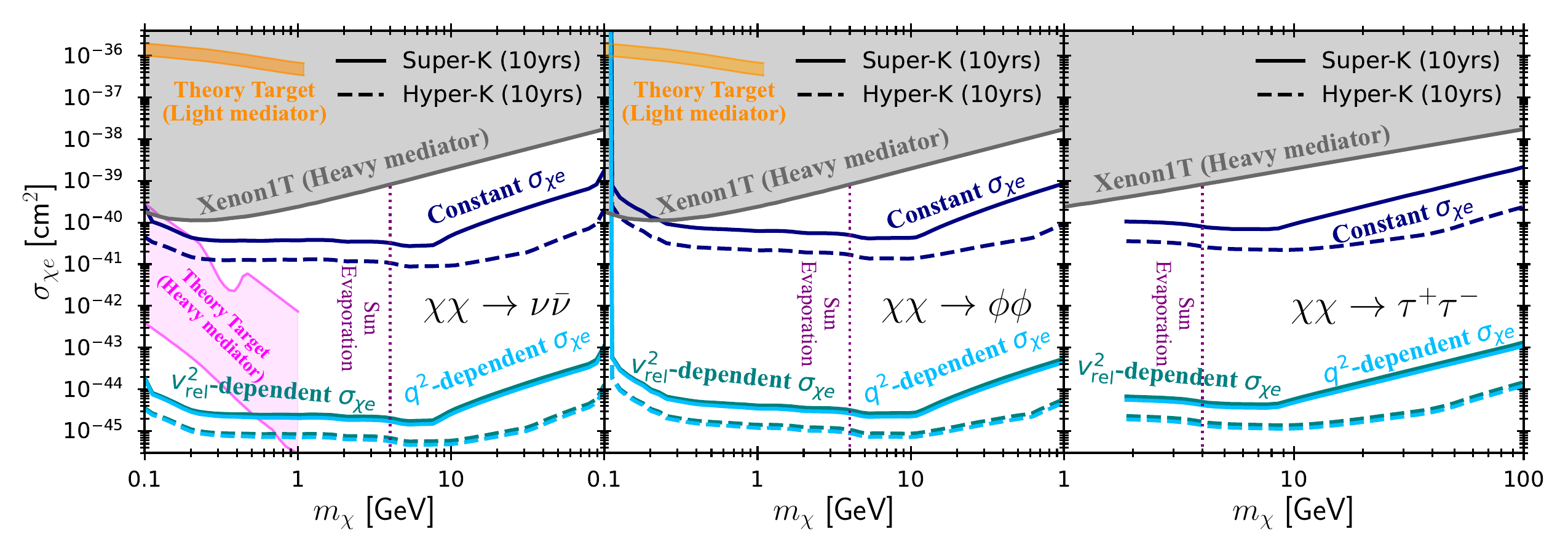}
\vspace{-0.4cm}
\caption{Super-K and Hyper-K $\sigma_{\chi e}$ bounds for different capture scenarios and annihilation channels: ({\bf Left}) $\chi\chi\to\nu\nu$, ({\bf Middle}) $\chi\chi\to\phi\phi$, ({\bf Right}) $\chi\chi\to\tau^{+}\tau^{-}$. The constant, velocity-dependent, and transfer momentum-dependent cross sections are navy, teal, and dodger blue, respectively. The theory target for sub-GeV dark matter with heavy-mediators are in magenta, while the theory target for light-mediator models is in orange~\cite{DDLimits, Essig:2022dfa}.}
\label{fig:othercapture}
\end{figure*}

\vspace{0.5cm}
\noindent \textbf{\emph{Results }}--- Figure~\ref{fig:firstpage} (bottom) shows the resulting constraint on the dark matter/electron cross-section as a function of the dark matter mass for dark matter masses between 0.1--100~GeV.  We consider three different leptophilic annihilation channels: The \mbox{$\chi\chi\to\nu \bar{\nu}$} and \mbox{$\chi\chi\to\phi\phi$} channels can push the limit down to $4\times10^{-41}$~cm$^{2}$ for masses below 10~GeV, while the \mbox{$\chi\chi\to \tau^{+}\tau^{-}$} channel provides the weakest limit of \mbox{$6\times 10^{-41}$~cm$^{2}$} for the same dark matter mass range. We note that for masses below 4~GeV, dark matter evaporation may significantly weaken the limits~\cite{Garani:2021feo}, though the detailed modeling of dark matter evaporation is complex~\cite{Acevedo:2023owd}. Furthermore, in the case of electron scattering, the evaporation mass can vary between a few hundred MeV up to 4~GeV, depending on the cut-off of the DM velocity distribution inside the Sun~\cite{Garani:2017jcj}. Therefore, we show the extended bounds and projections for $\chi\chi\to\nu\bar{\nu}$ and $\chi\chi\to\phi\phi$ channels down to 0.1~GeV, and for $\chi\chi\to\tau^{+}\tau^{-}$ down to 1.8~GeV.

Excitingly, these SuperK limits are at least 1 order of magnitude stronger than the leading terrestrial limits on leptophilic dark matter scattering by the Xenon Collaboration~\cite{XENON:2019gfn}.
We note that our results depend on the number of observed neutrinos (rather than their total energy), and thus the three annihilation final states that we test provide similar limits. Thus, any standard leptophilic dark matter model will be expected to have scattering constraints of approximately $10^{-40}$~cm$^{2}$ for dark matter masses between 4--100~GeV.

In Figure~\ref{fig:compare}, we show projections for a similar 10~yr observation using Hyper-K, finding that it provides an improvement of roughly one order of magnitude compared to Super-K limits. We compare our results to pervious studies of leptophilic dark matter capture and annihilation inside the Sun~\cite{Kopp:2009et, Maity:2023rez}, finding that our cross-section limits are an order of magnitude stronger with Super-K and two orders of magnitude stronger with Hyper-K, and also extend to lower dark matter masses.

In Figure~\ref{fig:othercapture} we calculate the dark matter scattering limits for alternative dark matter/electron capture scenarios, including constant, velocity-dependent, and transfer momentum-dependent cross sections.  We include the ``Theory target" parameter spaces for sub-GeV leptophilic dark matter, with either a heavy ~\cite{DDLimits, Essig:2022dfa, Essig:2015cda, Boehm:2003hm, Boehm:2003ha, Fayet:2004bw, Choi:2017zww, Lin:2011gj, Hochberg:2014kqa, Hochberg:2014dra, Kuflik:2015isi, Kuflik:2017iqs, DAgnolo:2018wcn} or a light mediator ~\cite{DDLimits, Essig:2022dfa, Chu:2011be, Dvorkin:2019zdi, Chang:2019xva}. We note that terrestrial direct detection constraints on light mediators are much less sensitive, only reaching  $\sim10^{-35}$~cm$^{2}$, which we show in Figure~\ref{fig:JupBound} of Appendix~\ref{appen:jupiter}~\cite{SENSEI:2020dpa, SENSEI:2024yyt, DAMIC-M:2023gxo, DAMIC-M:2023hgj, Essig:2017kqs, PandaX-II:2021nsg}.

Excitingly, the two dark matter capture scenarios with velocity- and momentum-dependent cross sections produce constraints that are up to 4--5 orders of magnitude stronger than the standard case of isotropic and velocity-independent cross sections. This makes our results in Figures~\ref{fig:firstpage}~and~\ref{fig:compare} the most conservative constraint compared to different models for dark matter capture. If the Solar evaporation barrier can be pushed down, these bounds can exclude the entire freeze-in theory target from 0.1--1~GeV. In the case of constant cross sections, Super-K and Hyper-K can only exclude (or potentially exclude) a small fraction of the heavy-mediator theory target, while the other two scenarios can exclude most the parameter space from 0.1--1~GeV. These results demonstrate the potential of celestial body neutrino searches to exclude theory targets for dark matter-electron scattering, which we leave for future work~\cite{Carlos_paper}.

\vspace{0.5cm}
\noindent \textbf{\emph{Conclusions }}--- We use Super-K observations to strongly constrain leptophilic dark matter capture in the Sun. We study the neutrino signals from all leptophilic annihilation channels, finding that annihilations to $\tau^{+}\tau^{-}$, light-mediators, and the direct annihilation to neutrinos provide the strongest constraints. We obtain world-leading limits on leptophilic dark matter scattering, reaching below 10$^{-40}$~cm$^{2}$ for dark matter masses between 4--20~GeV, and exceeding terrestrial direct detection searches by an order of magnitude between 4--100~GeV and project that Hyper-K data can improve these constraints by an order of magnitude.

By further extending our analysis down to a  dark matter mass of 0.1~GeV and considering different dark matter capture scenarios, we demonstrate the potential of the Sun and other celestial bodies to probe the theory parameter space of dark matter/electron scattering~\cite{DDLimits, Essig:2022dfa, Essig:2015cda, Boehm:2003hm, Boehm:2003ha, Fayet:2004bw, Choi:2017zww, Lin:2011gj, Hochberg:2014kqa, Hochberg:2014dra, Kuflik:2015isi, Kuflik:2017iqs, DAgnolo:2018wcn, Chu:2011be, Dvorkin:2019zdi, Chang:2019xva}. Notably, our work obtains larger capture rates compared to previous work on the topic, which we find to be due to the extra stopping power of high-momentum electrons near the solar core. This result significantly enhances the expected sensitivity of all solar searches for dark matter electron scattering. Our results demonstrate the potential of neutrino detectors to investigate leptophilic dark matter, which strengthens the motivation for future experiments such as JUNO~\cite{JUNO:2021tll} and DUNE~\cite{DUNE:2020ypp}.

\vspace{0.2cm}
\noindent \textbf{\emph{Acknowledgement }}--- We thank Carlos Blanco, Chau Thien Nhan, and Ludwig Neste for fruitful discussions; Bei Zhou and Shirley Li for help with the background modeling of Super-K and Hyper-K; and Chris Cappiello, Dhashin Krishna, Ranjan Laha, Tarak Nath Maity, Nirmal Raj, Akash Kumar Saha, and Rinchen Sherpa for comments on the manuscript. We thank Raghuveer Garani, Joachim Kopp, and Sergio Palomares-Ruiz for helping us cross-check our results. TTQN thanks Jorge Martin Camalich and the Instituto de Astrof\'{\i}sica de Canarias for their hospitality. TTQN, TL and PC are supported by the Swedish Research Council under contract 2022-04283. TL is also supported by the Swedish National Space Agency under contract 2023-00242. This article/publication is based on the work from COST Action COSMIC WISPers CA21106, supported by COST (European Cooperation in Science and Technology). This work made use of {\tt Numpy}~\cite{Harris_2020}, {\tt SciPy}~\cite{Virtanen:2019joe}, {\tt matplotlib}~\cite{HunterMatplotlib}, {\tt Jupyter}~\cite{2016ppap.book...87K}, as well as {\tt Webplotdigitizer}~\cite{Rohatgi2022}.

\clearpage
\newpage
\maketitle
\onecolumngrid
\begin{center}
\textbf{\large Super-Kamiokande Strongly Constrains Leptophilic Dark Matter Capture in the Sun}

\vspace{0.05in}
{ \it \large Supplemental Material}\\ 
\vspace{0.05in}
{Thong T.Q. Nguyen, Tim Linden, Pierluca Carenza, Axel Widmark}
\end{center}
\onecolumngrid
\setcounter{equation}{0}
\setcounter{figure}{0}
\setcounter{section}{0}
\setcounter{table}{0}
\setcounter{page}{1}
\makeatletter
\renewcommand{\theequation}{S\arabic{equation}}
\renewcommand{\thefigure}{S\arabic{figure}}
\renewcommand{\thetable}{S\arabic{table}} 

\tableofcontents

\section{Dark matter capture rate}
\label{appen:capt_rate}
Dark matter particles of mass $m_{\chi}$ and velocity (at infinity) $u_{\chi}$ can fall into the gravitational potential of celestial bodies, gaining velocities $w(r)=\sqrt{u_{\chi}^{2}+v_{\rm esc}^{2}(r)}$, which depend on the escape velocity $v_{\rm esc}(r)$ at a given position relative to the object, calculated as:
\begin{equation}
    v^{2}_{\rm esc}(r)=\int_r^{\infty} \frac{2G _{N}\Big[\int_0^{r'} 4\pi r''^{2}\rho(r''){\rm d}r'' \Big] }{r'^2} {\rm d}r',
    \label{eq:vesc}
\end{equation}
where $r'$ is the dummy variable corresponding to the dark matter position, and $r''$ is the second dummy variable corresponding to the mass density distribution of the celestial object. The stellar mass density at position $r''$ is given by $\rho$, and $G_{N}$ is the gravitational constant. 

Leptophilic dark matter can scatter with electrons inside the celestial object and lose kinetic energy. The celestial body will capture the dark matter particle if its velocity becomes smaller than the escape velocity. In the limit where dark matter particles only scatter with the celestial object once, the dark matter particles will be captured at a rate given 
\begin{align}
    C^{\rm weak}_{\star}=&\int_{0}^{R_{\star}}{\rm d}r4\pi r^{2}\int_{0}^{\infty}{\rm d}u_{\chi}\Big{(}\frac{\rho_{\chi}}{m_{\chi}}\Big{)}\frac{f_{v_{\odot}}(u_{\chi})}{u_{\chi}}\times\int_{0}^{v_{\rm esc}(r)}{\rm d}v\times w(r) R^{-}_{e}(w\to v),
    \label{eq:captweak}
\end{align}
\noindent where the dark matter density for celestial bodies inside the solar system is $\rho_{\chi}=0.3$~GeV/cm$^{3}$. Therefore, the dark matter velocity distribution is given by:

\begin{equation}
    f_{v_{\odot}}(u_{\chi})=\sqrt{\frac{3}{2\pi}}\frac{u_{\chi}}{v_{\odot}v_{d}}\Big{(}e^{-\frac{3(u_{\chi}-v_{\odot})^{2}}{2v_{d}^{2}}}-e^{-\frac{3(u_{\chi}+v_{\odot})^{2}}{2v_{d}^{2}}}\Big{)},
    \label{eq:fvdm}
\end{equation}

\noindent where the solar velocity with respect to the dark matter rest frame is $v_{\odot}=220$~km/s and the galactic velocity dispersion is $v_{d}=\sqrt{3/2}v_{\odot}\simeq 270$~km/s. 
While the capture rate in Eq.~\eqref{eq:captweak} is calculated by integrating over the entire dark matter velocity distribution (through the integration over $u_{\chi}$), the standard halo distribution is suppressed at velocities that exceed the galactic escape velocity, which is $v_{\rm esc}^{\rm gal}=533^{+54}_{-41}$~km/s at 90\%~CL~\cite{Piffl:2013mla}. Thus, in our analysis, we add a cutoff in the dark matter velocity distribution at $u^{\rm gal}_{\rm esc}=1000$~km/s to account for the effect of dark matter escape.

The scattering rate for a dark matter with initial velocity $\omega(r)$ that go down to below the celestial velocity at a given position inside the object, $v\leq v_{\rm esc}(r)$, is given by~\cite{Garani:2017jcj}

\begin{equation}
    R_{e}^{-}=\frac{2}{\sqrt{\pi}}\frac{n_{e}(r)}{u_{e}^{3}(r)}\int_{0}^{\infty}{\rm d}u u^{2}\int_{-1}^{1}{\rm d}\cos\theta \frac{{\rm d}\sigma_{\chi e}}{{\rm d}v}|\bm{w}-\bm{u}|e^{-u^{2}/u_{e}^{2}(r)},
\end{equation}
assuming the targeted electrons, with number density $n_{e}(r)$, have a relative scattering angle $\theta$, and velocity $u$ that follows the Maxwell-Boltzmann velocity distribution. The most probable speed of electron,
\begin{equation}
    u_{e}(r)=\sqrt{2T_{\star}(r)/m_{e}},
\end{equation}
is a function of the temperature profile $T_{\star}(r)$ of the celestial body. The scattering rate depends on the distribution of dark matter-electron cross section by its velocity after the scattering, ${\rm d}\sigma_{\chi e}/{\rm d}v$, which is usually considered in three different scenarios: (1) isotropic and velocity-independent cross sections, (2) isotropic but velocity-dependent cross sections, or (3) momentum dependent cross sections. We follow the detailed calculations of the capture rate for each case from Ref.~\cite{Garani:2017jcj}. We use the notation from Ref.~\cite{Gould:1987ju} to define these functions:
\begin{align}
    &\mu=\frac{m_{\chi}}{m_{e}},\quad \mu_{\pm}=\frac{\mu\pm 1}{2},\\
    &\chi(a,b)=\int_{a}^{b}e^{-y^{2}}{\rm d}y = \frac{\sqrt{\pi}}{2}\Big{[}{\rm erfc}(a)-{\rm erfc}(b)\Big{]},\\
    &\alpha_{\pm}=\frac{\mu_{+}\times v \pm \mu_{-}\times w}{u_{e}(r)}, \quad\beta_{\pm}=\frac{\mu_{-}\times v \pm \mu_{+}\times w}{u_{e}(r)}.
\end{align}
\noindent In the following subsections, we provide the specific formulae for each capture rate.

\subsection{Constant cross section}
\label{subappen:constant_sig}

For the usual case in the literature, the dark matter-electron cross section is isotropic and independent of velocity. The scattering rate is:
\begin{equation}
    R^{-}_{\rm const}(w\to v)=\frac{2}{\sqrt{\pi}}\frac{\mu^{2}_{+}}{\mu}\frac{v}{w}n_{e}(r)\sigma_{\chi e}\Big{[}\chi(-\alpha_{-}, \alpha_{+})+\chi(-\beta_{-}, \beta_{+})e^{\mu(w^{2}-v^{2})/u_{e}^{2}(r)}\Big{]}.
    \label{eq:Rminus_const}
\end{equation}
\noindent This assumption come with the case where the dark matter interaction with electron, in particular, the form factor, is a constant. This case is usually implies that dark matter interacts with electron by a heavy mediator, or contact interaction. In Figure~\ref{fig:firstpage}, using this scattering rate, we compare our results from the Sun and Super-K with the constraints from Xenon1T that assumes heavy mediator~\cite{XENON:2019gfn}.

\subsection{Cross section depends on relative velocity ($v_{\rm rel}^{2}$)}
\label{subappen:vrel_sig}

For the case the cross section is still isotropic, but depends on the initial relative velocity between the dark matter particle and the targeted electron. In this case the scattering rate is:

\begin{equation}
    \begin{split}
        R^{-}_{v^{2}_{\rm rel}}(w\to v)=\frac{2}{\sqrt{\pi}}\frac{\mu_{+}^{2}}{\mu}\frac{v}{w}n_{e}(r)\sigma_{\chi e}\Big{(}\frac{u_{e}(r)}{v_{0}}\Big{)}^{2}&\Big{[}\Big{(}\mu_{+}+\frac{1}{2}\Big{)}\Big{(}\frac{w-v}{u_{e}(r)}e^{-\alpha^{2}_{-}}-\frac{w+v}{u_{e}(r)}e^{-\alpha_{+}^{2}}\Big{)}\\
        &+\Big{(}\frac{w^{2}}{u_{e}^{2}(r)}+\frac{3}{2}+\frac{1}{\mu}\Big{)}\chi(-\alpha_{-}, \alpha_{+})\\
        &+\Big{(}\frac{v^{2}}{u_{e}^{2}(r)}+\frac{3}{2}+\frac{1}{\mu}\Big{)}\chi(-\beta_{-},\beta_{+})e^{\mu(w^{2}-v^{2})/u_{e}^{2}(r)}\Big{]}.
    \end{split}
    \label{eq:Rminus_vrel}
\end{equation}

\noindent Compared to the case of a constant cross section, this scattering rate has an enhancement factor $(u_{e}/v_{0})^{2}$ which is proportional to the square of the targeted electron velocity.  We follow Ref.~\cite{Garani:2017jcj} and choose the arbitrary reference velocity $v_{0}=v_{\odot}= 220$~km/s, which is applicable for celestial objects inside the Solar system.

\subsection{Cross section depends on transfer momentum ($q^{2}$)}
\label{subappen:q2_sig}

Finally, we consider the case where the cross section depends on both the relative velocity and the dark matter mass, which combine to determine the transfer momentum $q^{2}=m_{\chi}^{2}|{\bm w}-{\bm v}|^{2}$. The scattering rate is given by:

\begin{equation}
    \begin{split}
        R^{-}_{q^{2}}(w\to v)=\frac{8}{\sqrt{\pi}}\frac{\mu^{4}_{+}}{\mu^{2}}\frac{v}{w}n_{e}(r)\sigma_{\chi e}\Big{(}\frac{u_{e}(r)}{v_{0}}\Big{)}^{2}&\Big{[}\frac{w-v}{u_{e}(r)}e^{-\alpha^{2}_{-}}-\frac{w+v}{u_{e}(r)}e^{-\alpha_{+}^{2}}\\
        &+\Big{(}\frac{1}{2}\frac{w^{2}-v^{2}}{u_{e}^{2}(r)}+\frac{1}{\mu}\Big{)}\chi(-\alpha_{-},\alpha_{+})\\
        &-\Big{(}\frac{1}{2}\frac{w^{2}-v^{2}}{u_{e}^{2}(r)}+\frac{1}{\mu}\Big{)}\chi(-\beta_{-},\beta_{+})e^{\mu(w^{2}-v^{2})/u_{e}^{2}(r)}\Big{]}.
    \end{split}
    \label{eq:Rminus_q2}
\end{equation}

\noindent Similar to the velocity-dependent case, the scattering rate gains an enhancement related to the electron velocity. This case assumes a form factor for dark matter interaction with an electron that depends on the transfer momentum. This is usually considered in the case of presence of a light mediator~\cite{Essig:2022dfa, Chu:2011be, Dvorkin:2019zdi, Chang:2019xva}.

\subsection{Celestial Object Saturation Cross Section}
\label{subappen:geo_sig}

We note the capture derivation we utilize in the main text is only formally true in the limit where the dark matter cross-sections are extremely small, implying that the probability that a dark matter particle scatters twice off of the celestial body is infinitesimal. In this limit, the capture rate is linearly proportional to the dark matter-electron cross-section, $\sigma_{\chi e}$, as shown in Eqs.~\ref{eq:Rminus_const},~\ref{eq:Rminus_vrel},~\ref{eq:Rminus_q2}. However, as the cross-section increases, the capture probability must be modified to account for the possibility of multiple scattering interactions~\cite{Leane:2021ihh, Leane:2023woh, Ilie:2021iyh, Ilie:2024sos}. For very high cross-sections, the capture rate is saturated at a limit, which is called the geometric cross-section~\cite{Bernal:2012qh, Bottino:2002pd}:

\begin{equation}
    C^{\rm geo}_{\star} = \pi R^{2}_{\star}\Big{(}\frac{\rho_{\chi}}{m_{\chi}}\Big{)}\langle v_{0}\rangle\Big{(}1+\frac{3v_{\rm esc}^{2}(R_{\star})}{v_{d}^{2}}\Big{)}\xi(v_{\odot,v_{d}}),
\end{equation}
where the average speed in the DM rest frame is given by \mbox{$\langle v_{0}\rangle=\sqrt{8/(3\pi)}v_{d}$.} The factor $\xi(v_{\odot}, v_{d})$ takes into account the suppression due to the motion of the celestial body. In the solar system, this suppression factor is $\xi\simeq 0.81$.

Finally, in to produce a smooth transition between high dark matter-electron cross sections and the geometric capture rate limit, the final capture rate formula is usually approximated using a functional form:
\begin{equation}
    C_{\star}=C^{\rm weak}_{\star}\Big{(}1-e^{-C^{\rm geo}_{\star}/C^{\rm weak}_{\star}}\Big{)}.
    \label{eq:Cfinal}
\end{equation}

\begin{figure}[tbp!]
\centering
\includegraphics[width=1\columnwidth]{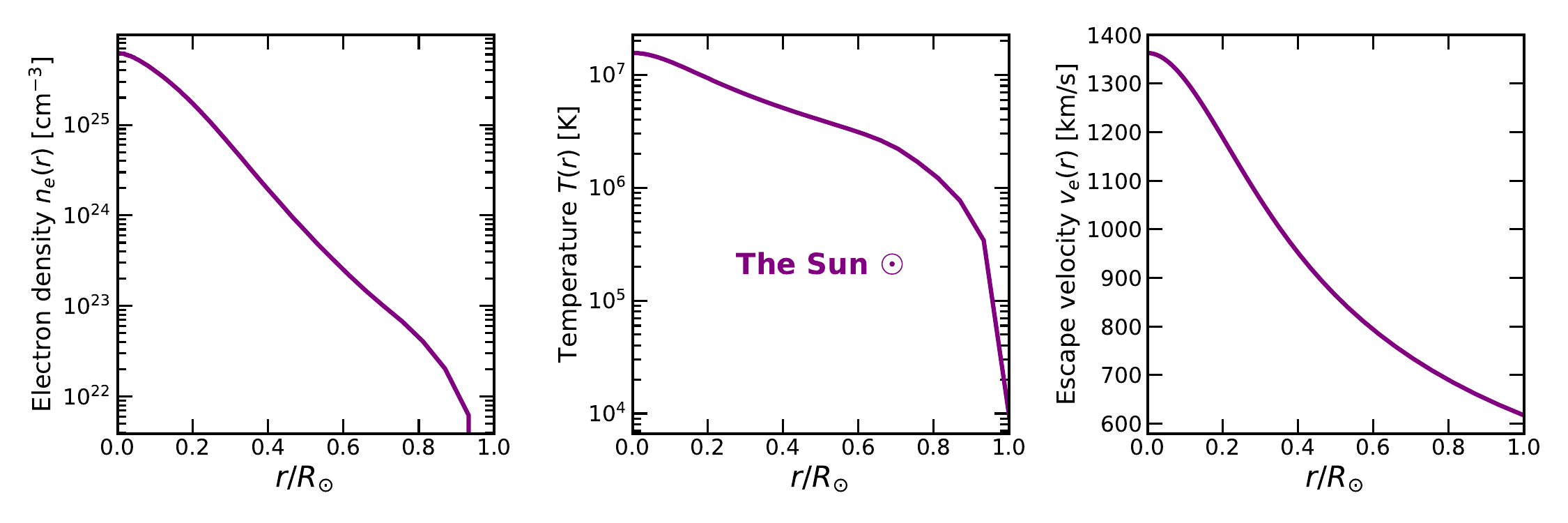}
\caption{A Standard Model of the \textcolor{violet}{Sun} as a function of its radius~\cite{Magg_2022}, including  {\bf Left:} the electron density, {\bf Middle:} the temperature profile, and {\bf Right:} the dark matter escape velocity.}
\label{fig:profile}
\end{figure}

\subsection{Dark Matter Capture in the Sun and Jupiter}
\label{subappen:sun_and_jupiter}
\begin{figure}[tbp!]
\centering
\includegraphics[width=1\columnwidth]{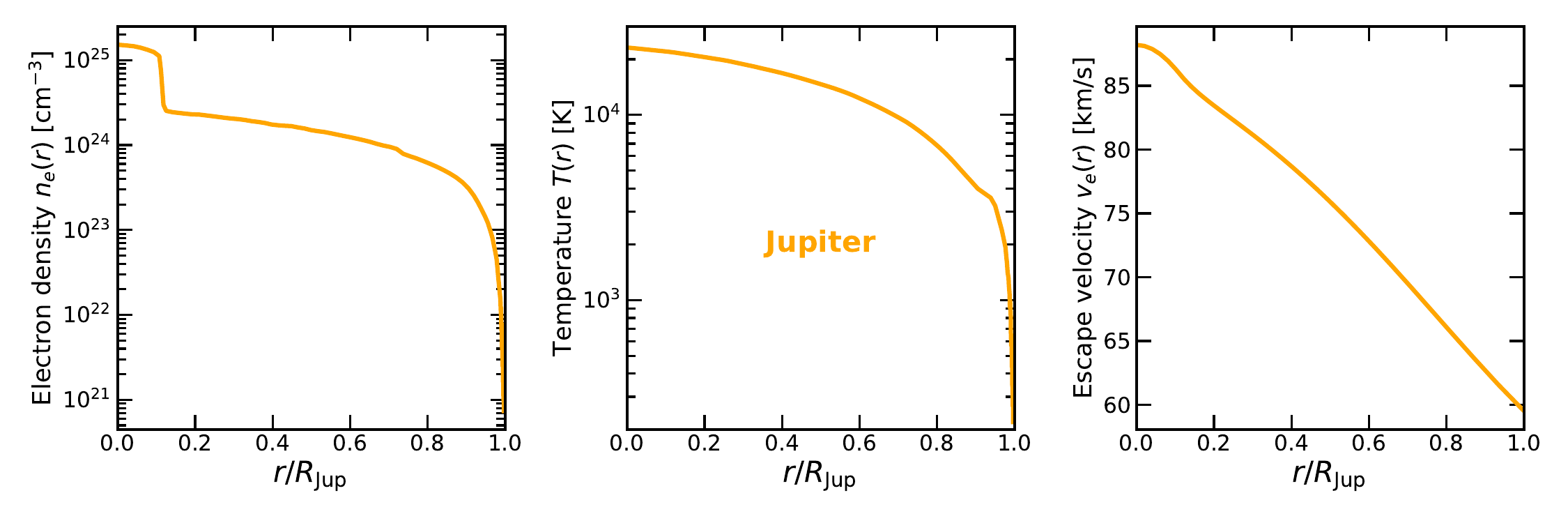}
\caption{Similar to Figure~\ref{fig:profile}, but for \textcolor{orange}{Jupiter}: We use the Jovian J11-4a model in Ref.~\cite{French_2012}.}
\label{fig:JupProfile}
\end{figure}

In this appendix, we discuss the contribution of the physical parameters of two solar system bodies that have been studied in the context of dark matter capture: the Sun (following the same approach as the main text)~\cite{Acevedo:2020gro, Berlin:2024lwe, Chauhan:2023zuf, Chu:2024gpe, Garani:2017jcj, Maity:2023rez, HAWC:2022khj, Nisa:2019mpb, HAWC:2018szf, IceCube:2021xzo, Leane:2017vag, Bell:2021esh, Bell:2011sn, Bell:2021pyy, Bell:2012dk, Niblaeus:2019gjk, Edsjo:2017kjk, Bose:2021cou, IceCube:2016yoy, Feng:2016ijc,Kouvaris:2007ay, Kouvaris:2015nsa, Kouvaris:2016ltf, Widmark:2017yvd, Catena:2016ckl}, as well as Jupiter~\cite{Linden:2024uph, Blanco:2023qgi, Blanco:2024lqw, Leane:2021tjj, Li:2022wix, Yan:2023kdg, Croon:2023bmu}. In Figures~\ref{fig:profile}~and~\ref{fig:JupProfile} we plot the electron density, electron temperature, and dark matter escape velocity for each object. For the dark matter escape velocities we follow Eq.~\ref{eq:vesc}.

\begin{figure}[tbp!]
\centering
\includegraphics[width=1\columnwidth]{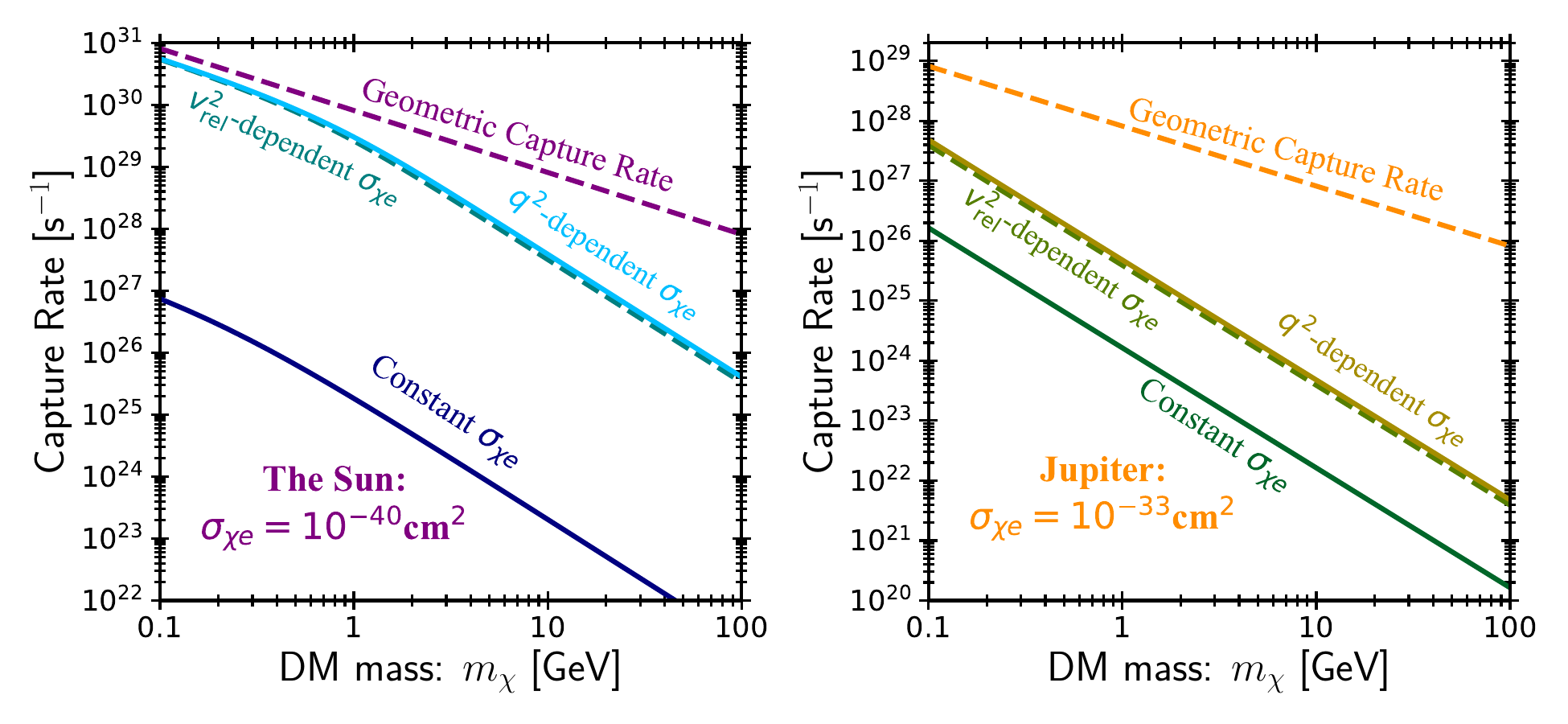}
\caption{The leptophilic dark matter capture rate, assuming electrons follow the Maxwell-Boltzmann distribution: ({\bf Left}) The Sun with $\sigma_{\chi e}=10^{-40}$~cm$^{2}$, with constant cross section (navy blue), velocity-dependent cross section (teal), and momentum-dependent (dodger blue) cross section. ({\bf Right}) Jupiter with $\sigma_{\chi e}=10^{-33}$~cm$^{2}$, with constant cross section (dark green), velocity-dependent (olive) cross section, and momentum dependent (dark goldenrod) cross section. The two geometric capture rates for the Sun and Jupiter are in purple and orange respectively.}
\label{fig:CaptRate}
\end{figure}

The Sun is the most well-known and closest star, Therefore it is a promising laboratory to probe new physics. The Sun is a main sequence star with a mass $M_{\odot}=1.98\times 10^{33}$~g, a radius $R_{\odot}= 6.96 \times 10^{10}$~cm, a core temperature of $T\sim1.5\times10^7$~K and a core density of $\sim 150~{\rm g}~{\rm cm}^{-3}$. The chemical composition of the Sun varies from the innermost core region, which is composed of $\sim 36\%$ of hydrogen ($^{1}$H), and $\sim 62\%$ helium ($^{4}$He); to an external envelope composed of $\sim 76\%$ of $^{1}$H and $\sim 23\%$ of $^{4}$He. In this work, we consider the Standard Solar Model from Ref.~\cite{Magg_2022}.

Jupiter is the largest and most massive planet in the Solar system, therefore it is well-suited to be used as a natural DM detector. Jupiter has a mass of $M_{\rm jup}=1.9\times10^{27}$~kg and a radius of $R_{\rm jup}=6.99\times10^{9}$~cm. On average, the Jupiter helium-to-hydrogen mass ratio is $Y=0.275$, while the outer atmosphere has $Y= 0.238$ in agreement with spectroscopic measurements. We consider the Jovian J11-4a model from Ref.~\cite{French_2012}.

Assuming all the electrons inside these celestial objects follow the Maxwell-Boltzmann distribution, we calculate the dark matter capture rate for the Sun and Jupiter for all three capture scenarios discussed in Appendix~\ref{appen:capt_rate}. We show the results of the capture rate for these objects with some benchmark cross sections in Figure~\ref{fig:CaptRate}. We rescale our results for the capture rate with the geometric limits using Eq.~\ref{eq:Cfinal}.

Comparing our results between the standard isotropic and velocity-independent cross section against our two alternative models, we find that velocity- or momentum-dependent capture rates can enhance solar dark matter capture by 3--4 orders of magnitude, while for Jupiter the enhancement is around 1 order of magnitude. This is due to the fact that the Solar temperature is about 3 orders of magnitude higher than for Jupiter, which increases the target velocity enhancement as $(u_{e}/v_{0})^{2}$. Our results in Figure~\ref{fig:CaptRate} also demonstrate that for different dark matter capture schemes, we have different dark matter saturation cross section.

In the main text, we have adopted a default model of isotropic and velocity-independent capture, which provides the weakest limit of the three scenarios. We note that this means, for the purposes of our comparison with direct detection - we have been extremely conservative in claiming a single order of magnitude improvement. In many well-motivated scenarios, the constraints from solar dark matter capture exceed terrestrial direct detection by 4--5 orders of magnitude. The advantage over direct detection can be enhanced even further, if the scattering interaction moves through a light mediator, which significantly decreases the sensitivity of terrestrial experiments.

\newpage

\section{Comparison of Capture Rate Calculations with Previous Work -- \\ On the Zero Temperature Approximation}
\label{app:gould}

After the initial submission of this manuscript, it was pointed out by Ref.~\cite{Krishna:2025ncv} that the capture rates calculated in this work exceed those calculated by Ref.~\cite{Garani:2017jcj} by approximately a factor of three at low dark matter masses, and by a factor of up to seven for high dark matter masses.\footnote{What follows in the remainder of this abstract is due to significant work in understanding the difference between this work and the initial version of the work in Ref.~\cite{Krishna:2025ncv} (v1), Ref.~\cite{Garani:2017jcj} and Ref.~\cite{Kopp:2009et}. We thank the authors of these manuscripts for their work in exchanging calculations and code in order to crystallize the key scientific issues which affected previous work on the topic.} In this section, we explain the difference between these results, and show that our original results are correct because they account for the fact the Solar electron temperature has a sizable impact on the kinematics of the dark matter scattering cross-section. The results of our paper remain unchanged from the initial version of the manuscript. The enhancement of the dark matter capture rate demonstrated below also applies to Refs.~\cite{Garani:2017jcj} and ~\cite{Krishna:2025ncv}.

\begin{figure}[t]
    \centering
    \includegraphics[width=1\linewidth]{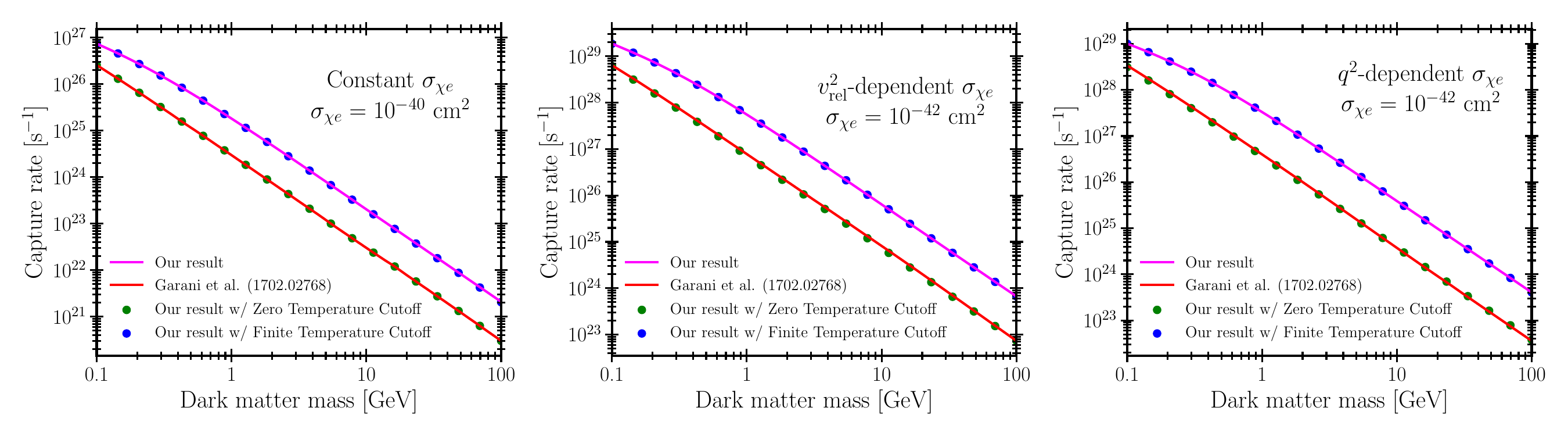}
    \caption{The total solar capture rate as a function of the dark matter mass and set at fixed cross sections of $\sigma_{\chi e} = 10^{-40}$~cm$^{2}$ and $\sigma_{\chi e}=10^{-42}$~cm$^{2}$ from our current work (magenta lines), and that of Garani et al.~\cite{Garani:2017jcj} (red lines). Results are shown for constant cross-sections (left), velocity-dependent cross-sections (center) and momentum-dependent cross-sections (right). Our capture rate is larger by a factor of three for low dark matter masses and up to seven for high dark matter masses. Modifying our results by adding a kinematic cutoff that was based on a zero-temperature electron population (Equation~\ref{eq:kinematic_cutoff_gould}) the green dotted points, which agrees with the result of Ref.~\cite{Garani:2017jcj}, while adding an (unnecessary) kinematic cutoff corresponding to the modified high-temperature electron population (Equation~\ref{eq:kinematics}) leads to the blue dotted points, and does not affect our result.}
    \label{fig:gouldcutoff}
\end{figure}

The key difference between our results and Ref.~\cite{Krishna:2025ncv} corresponds to an incorrect kinematic cutoff that has been added into the dark matter/electron scattering cross-section. This cutoff sets the total scattering rate to be zero in any regime where a zero-temperature electron would be unable to successfully capture the dark matter particle. In discussions with the authors of Ref.~\cite{Krishna:2025ncv}, we have confirmed that this is the primary difference between our results, and can confirm that our analyses reproduce each-other's capture rates when this kinematic cutoff is added or removed. To further confirm that adding this term allows us to reproduce previous work, in Figure~\ref{fig:gouldcutoff} we show a comparison between our capture rate and that of Ref.~\cite{Garani:2017jcj}, showing that the arbitrary addition of this kinematic cutoff reduces our capture rates to identically match the solution in Ref.~\cite{Garani:2017jcj}.

This zero-temperature kinematic cutoff was originally employed in the calculation of dark matter scattering with the Earth in Gould 1987~\cite{Gould:1987ir}, in Equations 2.10 -- 2.13. Gould also employs the zero-temperature approximation to simplify the expression from Equation 7.1 -- 7.5, which is confusingly called the ``finite temperature evaluation", though it is no longer accurate in the case of finite-temperature electrons. Notably, however, this term does not appear in the simultaneously published Gould 1987 paper for dark matter scattering in the Sun~\cite{Gould:1987ju}. 

While the zero-temperature approximation is very accurate for the 10$^4$~K temperatures in the Earth core, it produces sizable errors for the 10$^7$~K temperatures near the solar core, which correspond to the average electron moving at v$_{th}$ = $\sqrt{\frac{k_bT}{m_e}} \sim$~0.1~c. The high-velocity of these electrons gives them extra stopping power in head-on collisions with more massive dark matter particles~\footnote{It is also worth noting that Refs.~\cite{Gould:1987ir, Gould:1987ju} were primarily concerned with nuclear scattering, where the effect of the Earth temperature on the scattering cross-section is even more negligible, justifying their zero-temperature approximation.}

It is important to recognize the difference between the ``average stopping power" of an electron and the ``maximum stopping power" of an electron in a dark matter collision. The addition of a kinematic cutoff (given by a Heaviside function in the capture rate) is meant to represent the second case. In scenarios where a zero-temperature electron can slow down a dark matter particle below its escape velocity (for example at low dark matter masses), the total capture rate is maximized when all electrons are at zero temperature. Higher temperatures will decrease the dark matter capture rate by producing an electron sub-population that hits the dark matter particle from behind and further accelerates it. 

However, in a scenario where a zero-temperature electron \emph{cannot} slow down a dark matter particle to below its escape velocity (which is often the case for GeV-scale dark matter scattering off electrons in the Sun), the capture rate is largest when the electrons are at high temperature, and a sub-population of the electrons have the momentum necessary to slow down the dark matter particle in head-on collisions. Introducing a kinematic cutoff as soon as a zero-temperature electron cannot stop the dark matter particle ignores this large sub-population of electrons, which dominate the total capture rate for high-mass dark matter.

For clarity, we reproduce the calculation of the kinematic cutoff of Ref.~\cite{Gould:1987ir} below. For a dark matter mass m$_\chi$ and electron mass m$_e$, we first define the reduced masses:

\begin{equation}
    \mu = \frac{m_\chi}{m_e},
\end{equation}

and:

\begin{equation}
    \mu_{\pm} = \frac{\mu \pm 1}{2}.
\end{equation}

In a scenario where the electron has no momentum in the Sun's reference frame, simple kinematics then tells us that the kinetic energy lost by the dark matter particle in the reference frame of the Sun is then uniformly distributed over the range:

\begin{equation}
    0 \leq \frac{\Delta E}{E} \leq \frac{\mu}{\mu_+^2}.
\end{equation}

Assuming the dark matter particle stars at velocity $w(r)$ where it is unbound, and must scatter to a velocity smaller than $v_{\rm esc}(r)$ where it is bound to the Sun, we must have:

\begin{equation}
    \frac{\Delta E}{E}\Big{|}_{\rm max}^{\rm zero-temp} \geq \frac{w^2 - v_{esc}^2}{w^{2}} = \frac{u_{\chi}^2}{w^{2}},
\end{equation}

\noindent by combining these two conditions we are left with a Kronecker delta function which can be applied to the scattering rate to ensure that the scattering results in a dark matter particle that is bound to the Sun. 

\begin{equation}
    \label{eq:kinematic_cutoff_gould}
    C_\star^{{\rm weak}} \propto \theta\left(\frac{\mu}{\mu_+^2} - \frac{u_{\chi}^{2}}{w^2}\right).
\end{equation}

In the case where the electron has a high-momentum, this calculation clearly no longer applies, because it sets the capture rate to 0 based on a calculation that includes only the electron mass. If the electron has a significant momentum that is anti-parallel to the dark matter particle, conservation of momentum requires that it has additional stopping power capable of decreasing the dark matter velocity below v$_{esc}$, leading to dark matter capture. For an electron with an initial velocity $-u_e$ (in the opposite direction of the dark matter), which collides with a dark matter particle with velocity $w$, one can calculate the maximum energy loss as:

\begin{equation}
    \label{eq:kinematics}
    \frac{\Delta E}{E}\Big{|}_{\rm max}^{\rm finite-temp} = \frac{4m_{\chi}m_{e}}{(m_{\chi}+m_{e})^{2}}\frac{w+u_{e}}{w^{2}}\left(w-\frac{m_{e}}{m_{\chi}}u_{e}\right) = \frac{\mu}{\mu_{+}^{2}} \frac{w+u_{e}}{w^{2}}\left(w-\frac{m_{e}}{m_{\chi}}u_{e}\right).
\end{equation}

Compared to the kinematic cutoff in Eq.~\ref{eq:kinematic_cutoff_gould}, this maximum energy loss extends to significantly higher values in the case that $u_e$ is large, corresponding to an electron that is moving rapidly in the opposite direction of the dark matter particle. 

However, we note that Equation~\ref{eq:kinematics} should also not be translated into a simple theta-function cutoff for the dark matter capture rate (although for the solar temperatures, dark matter masses, and dark matter velocities in our study, applying it artificially to our calculated capture rate has no effect on our results, as shown in Figure~\ref{fig:gouldcutoff}). There are two reasons why dark matter scattering off a high temperature-target cannot be limited by a single Kronecker-delta function cutoff: (1) there will be high-energy electrons up to arbitrarily high energies stemming from the tails of the Maxwell-Boltzmann distribution of the electron population, (2) the distribution of energy losses below the high-energy cutoff is no longer uniform, and is not represented by a Kronecker-delta function. 

Fortunately, the original work of Gould (1987)~\cite{Gould:1987ju} already correctly accounts for the full calculation of the capture rates for a Maxwellian distribution of dark matter particles scattering off of a Maxwellian distribution of standard model targets. The kinematics of the scatter, which enforces that the final velocity of the dark matter particle ($v$) falls below the escape velocity of the Sun is already correctly accounted for in Equation~\ref{eq:captweak}, by calculating the scattering probability $R_e^-(w \rightarrow v)$ and enforcing that the final value of $v$ falls below the escape velocity $v_{{\rm esc}}(r)$ by setting this as the maximum limit of integration in $dv$. Thus, we conclude that Equation~\ref{eq:captweak}, as discussed in the initial draft of this paper, provides the correct formalism for scattering off of high-temperature electrons in the Sun.

The comparison with Ref.~\cite{Garani:2017jcj} is somewhat trickier, because their work utilized the full finite-temperature calculation for electron scattering. However, recent discussions have identified a numerical artifact in the integrator utilized by Ref.~\cite{Garani:2017jcj}, which the authors of that paper have recently corrected. When this numerical issue is corrected, the capture rate calculations of Ref.~\cite{Garani:2017jcj} match the results shown in this paper.

We note that applying our new results to several previous papers in the literature {\it e.g.},~\cite{Kopp:2009et, Garani:2017jcj, Maity:2023rez}) will enhance the capture rates in those calculations by factors of between 3--7, generally strengthening the conclusions of their work. Additionally, these results strengthen our projection for the sensitivity of future observatories (e.g., DUNE and Hyper-K) to dark matter/electron scattering. 
\color{black}

\section{Dark Matter Annihilation Rate Inside Celestial Bodies}
\label{appen:annihilationrate}
While each capture event adds a single dark matter particle the star, the dark matter density can also decrease due to evaporation or annihilation. The differential equation describing the number of dark matter particles in the star as a function of time is given by:
\begin{equation}
    \frac{{\rm d}N_{\chi}(t)}{{\rm d}t}=C_{\star}-C_{E}N_{\chi}(t)-C_{A}N_{\chi}^{2}(t),
\end{equation}
where $C_{E}$ is the evaporation rate, which depends on both the dark matter mass and the properties of the stellar object~\cite{Garani:2021feo}. The term $C_{A}$ is proportional to the annihilation rate of captured dark matter as, which can be approximated as:
\begin{equation}
    C_{A}=\langle \sigma_{\chi\chi} v\rangle/V_{\rm eff},
\end{equation}
where $V_{\rm eff}$ is the effective volume in which dark matter annihilation can take place, and serves as a proxy for calculating the effective density of dark matter within the star~\cite{Bhattacharjee:2022lts}. The number of captured dark matter particles within the stellar volume is then given by: 
\begin{equation}
    N_{\chi}(t)=\sqrt{\frac{C_{\star}}{C_{A}}}\tanh\frac{t}{t_{\rm eq}},
\end{equation}
where $t_{\rm eq}=1/\sqrt{C_{\star} C_{A} }$ is the equilibrium timescale of the stellar object. The annihilation rate of captured dark matter is:
\begin{equation}
    \Gamma_{\chi}=\frac{N_{\chi}(t)^{2}}{4V_{\rm eff}}\langle \sigma_{\chi\chi} v\rangle.
\end{equation}

For the majority of reasonable capture and annihilation cross-sections (those that are reasonably detectable) this equilibrium timescale is short compared to the age of the celestial object~\cite{Leane:2021ihh, Leane:2022hkk, Linden:2024uph, Maity:2023rez, Robles:2024tdh}. In this case, the annihilation rate simplifies significantly to:

\begin{equation}
    \Gamma_{\chi}=\frac{C_{\star}}{2},
\end{equation}
where the factor of two corresponds to the fact that each annihilation removes two dark matter particles from the object. 

\begin{figure}[t]
    \centering
    \includegraphics[width=0.5\linewidth]{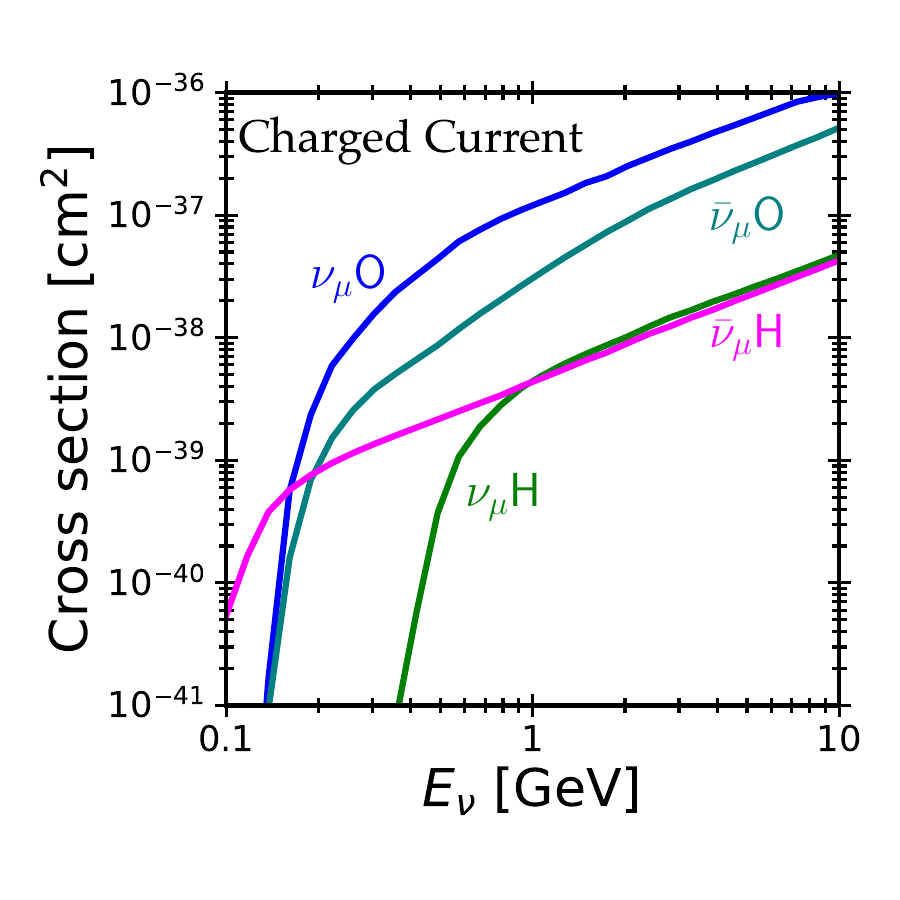}
    \caption{Charged Current interaction cross sections between muon neutrinos and anti-neutrino with Hydrogen and Oxygen inside water molecule.}
    \label{fig:H2O}
\end{figure}

\section{Neutrinos interaction inside the Sun and Water Cherenkov Detectors}
\label{appen:neutrino_int}

\subsection{Neutrino cross section and attenuation}
\label{subappen:neutrino_sigma}

Dark matter can annihilate and produce neutrinos through multiple channels such as direct annihilation, secondary decays or weak bremsstrahlung. These neutrinos can escape the stars due to its weak interaction~\cite{Maity:2023rez, Robles:2024tdh, Bose:2024wsh, Bose:2021yhz, Chauhan:2023zuf}. In principle however, these neutrinos still can interact with molecules inside stars and be absorbed through charged current weak interactions. This interaction becomes dominant when the neutrino energy approaches the weak scale, allowing the $W-$boson exchange to be on-shell.

In Figure~\ref{fig:H2O}, we show the cross-section results of muon neutrinos with hydrogen and oxygen molecules, taken from Ref.~\cite{Zhou:2023mou}. These results are generated using \texttt{GENIE}, which accounts for the comprehensive treatment of neutrino-nucleon and neutrino-nucleus interaction vertices, including nuclear effects, hadronization processes, final-state interactions, de-excitation of the residual nucleus, and additional relevant phenomena~\cite{Andreopoulos:2009rq}.

\begin{figure}[t]
    \centering
    \includegraphics[width=0.5\linewidth]{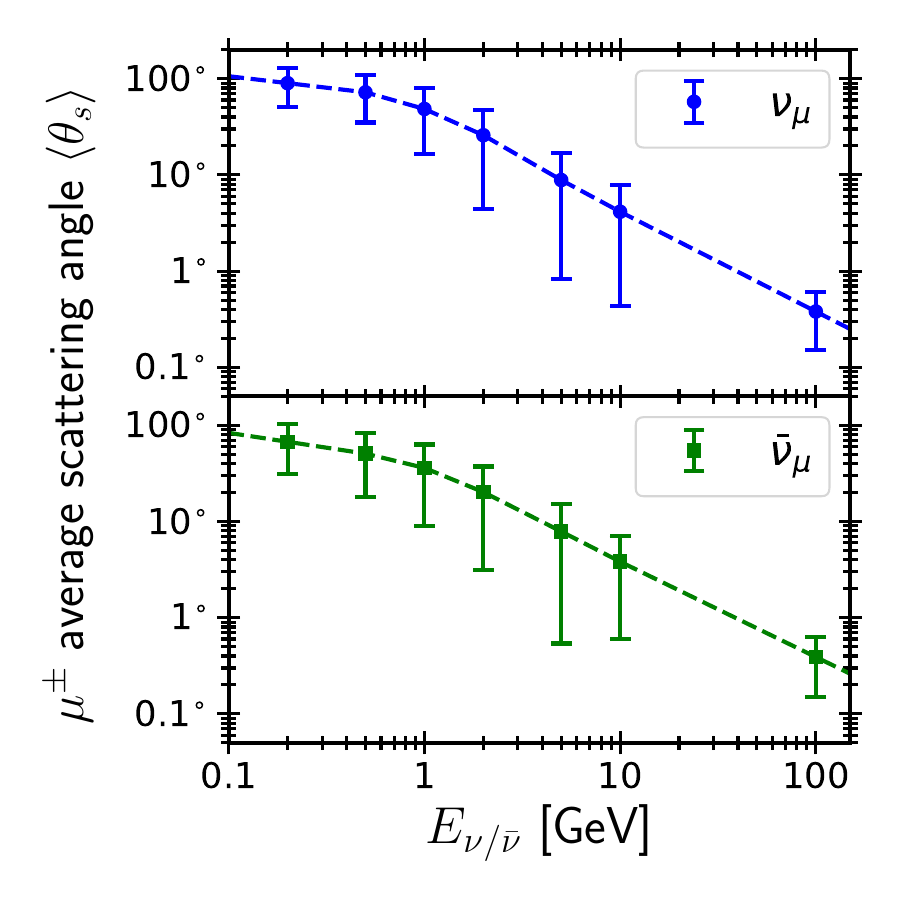}
    \caption{Average scattering angle of the neutrino (blue) and anti-neutrino (green) as a function of the signal energy for Water Cherenkov detectors. Interpolation functions for these angular resolutions are shown with dashed lines.}
    \label{fig:anlge}
\end{figure}

We apply this cross-section to calculate neutrino propagation through the Sun, assuming 100\% hydrogen composition (increasing the Helium abundance only slightly changes these results). Using the travel distance equal to the Sun's radius, the survival probability of neutrinos is given by:

\begin{equation}
    P_{\rm surv}^{\nu}(E_{\nu})=\exp\left[-\int_{0}^{R_{\odot}}{\rm d}r \sigma_{\nu H}(E_{\nu})n_{H}(r) \right],
    \label{eq:atten}
\end{equation}

\noindent which have the minimum value around 96\% at 100~GeV neutrino energy.

We use the cross-section results for muon neutrinos/anti-neutrinos with water molecules to calculate the number of events inside Water Cherenkov detectors, such as Super-K and Hyper-K.

\subsection{Neutrino angular resolution and background cuts for Super-K}
\label{subappen:neutrino_angle}

The main background in our analysis comes from the atmospheric muon neutrino background, shown in Figure~\ref{fig:firstpage} as the purple line~\cite{Honda:2011nf}. However, this background corresponds to a full-sky observation, which corresponds to approximately $\Omega_{\rm sky}^{2} \sim 40000^{\circ^{2}}$. Super-K, on the other hand, can reduce this background through its angular resolution, especially for muon neutrinos, due to their distinguishable tracks inside the detector after scattering with water molecules.

We use the average scattering angle measurement by the Super-K detector from Refs.~\cite{Konishi:2010mv, Konishi:2011sc} for muon neutrinos/anti-neutrinos, which we show in Figure~\ref{fig:anlge}. We use interpolation functions fit to both measurements, shown as dashed lines in our analysis. We also take into account the detector angular resolution for muon neutrinos inside the detector, which is around 4--5~degrees for 1--5~GeV muons~\cite{Galkin:2008qe} and around 1~degree above TeV muon energies~\cite{Super-Kamiokande:2007uxr}. Defining this muon angular resolution as $\langle \theta_{\mu} \rangle$, we calculate the total angular resolution for the atmospheric background of the Super-K detector as:

\begin{equation}
    \Omega_{\rm bg}^{2}\simeq \langle\theta_{s}\rangle^{2} + \langle \theta_{\mu} \rangle^{2}.
\end{equation}

We use this squared neutrino angular resolution to calculate the reduced atmospheric background, which we show as the dashed magenta line in Figure~\ref{fig:firstpage}. Notably, our effort to reduce the atmospheric background and investigate the neutrino signal from solar objects is similar to Ref.~\cite{Robles:2024tdh}. However, Ref.~\cite{Robles:2024tdh} uses a constant angular resolution of 20$^{\circ}$, whereas the actual angular resolution decreases from $\sim$90$^{\circ}$ to $\sim$2$^{\circ}$ over the energy range of 0.1--100~GeV. Our analysis also suggests that the results of Super-K and Hyper-K sensitivities on the dark matter-nucleon cross section may shift slightly due to angular resolution corrections.

\section{Bounds on dark matter-electron cross section with Jupiter}
\label{appen:jupiter}

\begin{figure}[tbp!]
\centering
\includegraphics[width=1\columnwidth]{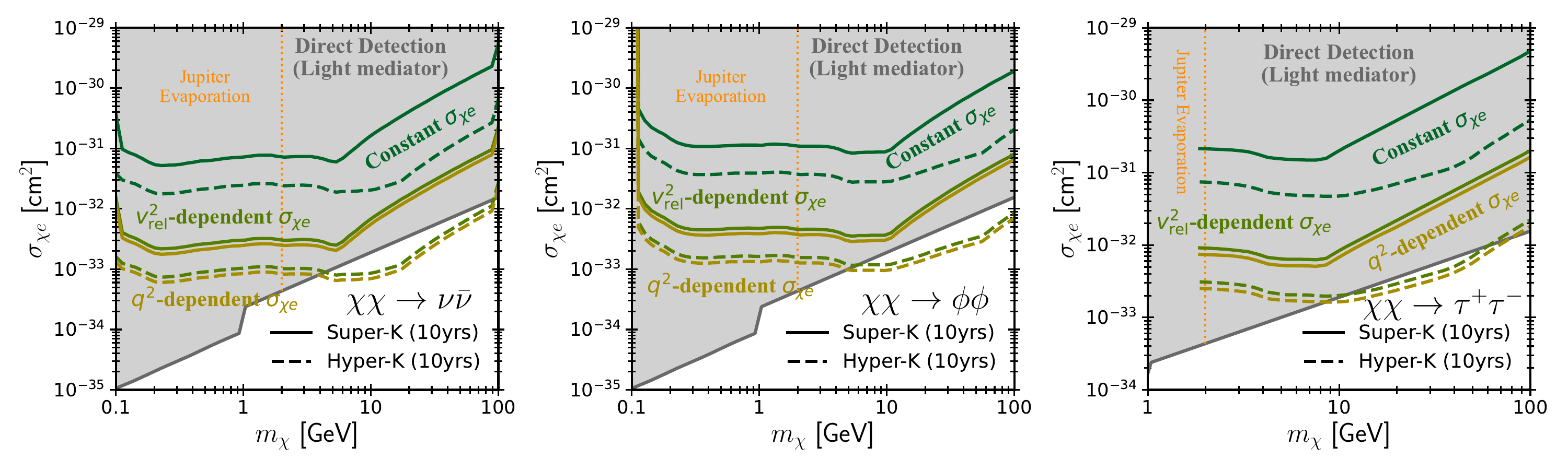}
\caption{Jupiter constraints on the dark matter-electron cross section for all dark matter capture scenarios: constant cross-section (dark green), relative velocity-dependent cross-section (dark goldenrod), and momentum-dependent cross-section (olive). Super-K results are shown with solid lines, while projections from Hyper-K are represented with dashed lines. These results are for different annihilation channels: $\chi\chi \to \nu\bar{\nu}$ ({\bf left}), $\chi\chi \to \phi\phi$ ({\bf middle}), and $\chi\chi \to \tau^{+}\tau^{-}$ ({\bf right}). The Jupiter evaporation barrier is indicated by the dotted orange line. Direct detection limits, shown in gray, are taken from Ref.~\cite{DDLimits}.}
\label{fig:JupBound}
\end{figure}

Using the same background cuts for atmospheric neutrinos and the analysis method for the Sun, we also calculate cross-section bounds for Jupiter. We adopt the average distance between the Sun and Jupiter as $623.32\times 10^{6}$~km. For the attenuation effect, we assume 100\% hydrogen composition, similar to Refs.~\cite{Leane:2022hkk}, for the least absorption factor, though this may not be true for the real properties of Jupiter.

We show all the Super-K dark matter-electron cross-section bounds from Jupiter for all leptophilic dark matter annihilation signals in Figure~\ref{fig:JupBound}. We also include projections for Hyper-K with the same 10-year timescale. These bounds have an evaporation barrier around 2~GeV for Jupiter. We compare them with the best limits from direct detection for light mediators from Refs.~\cite{Carew:2023qrj, Essig:2022dfa, DDLimits}, which combine results from SENSEI~\cite{SENSEI:2020dpa, SENSEI:2024yyt}, DAMIC~\cite{DAMIC-M:2023gxo, DAMIC-M:2023hgj}, Xenon10~\cite{Essig:2017kqs}, PandaX~\cite{PandaX-II:2021nsg}, and DarkSide~\cite{DarkSide:2022knj}. We note that Jupiter bounds using Super-K for all dark matter capture scenarios are entirely excluded by direct detection methods. Only projections from Hyper-K with a 10-year timescale are slightly better for velocity- and momentum-dependent cross section dark matter capture compared to current limits. We note that Jupiter cannot produce any bounds for heavy-mediator direct detection scenarios that exceed the bounds from Solar observations. Notably, we calculate all these results for Jupiter under the assumption that Jupiter is composed of 100\% hydrogen and that the electrons inside this object follow a Maxwell-Boltzmann distribution. In reality, Jupiter has more complex properties, which may be treated properly in the dark matter capture rate in future work~\cite{Carlos_paper}.

\bibliography{ref}

\end{document}